# On the depth resolution of transmission Kikuchi diffraction (TKD) analysis

Junliang Liu*, Sergio Lozano-Perez, Angus J. Wilkinson, Chris R.M. Grovenor

Department of Materials, University of Oxford, Parks Road, OX1 3PH, United Kingdom

*Corresponding author: bryanchina@hotmail.com

**Abstract**

In this paper, we have analysed the depth resolution that can be achieved by 'on-axis' transmission Kikuchi diffraction (TKD) using a Zr-Nb alloy. The results indicate that the signals contributing to detectable Kikuchi bands originate from a depth of approximately the mean free path of thermal diffuse scattering ($\lambda_{TDS}$) from the bottom surface of a thin foil sample. This existing surface sensitivity can thus lead to the observation of different grain structures when opposite sides of a nano-crystalline foil are facing the incident electron beam. These results also provide a guideline for the ideal sample thickness for TKD analysis of $\leq 6\lambda_{TDS}$, or 21 times the elastic scattering mean free path ($\lambda_{MFP}$) for samples of high crystal symmetry. For samples of lower symmetry, a smaller thickness $\leq 3\lambda_{TDS}$, or $\leq 10\lambda_{MFP}$ is suggested.

## 1. Introduction

Transmission Kikuchi diffraction (TKD) [1], also sometimes called transmission-electron backscatter diffraction (t-EBSD), is a relatively new method for orientation mapping of the microstructures of thin transmission electron microscopy specimens in the scanning electron microscope and has been widely utilized in the characterisation of nano-crystalline materials, including oxides [2], superconductors [3] and metallic alloys [4]. By using an electron-transparent thin TEM foil instead of a bulk sample, the probed sample volume using a focused, transmitted electron beam is much smaller and the lateral resolution can be improved to ~3 nm [5].

Compared to the conventional EBSD and 'off-axis' TKD configurations, the new 'on-axis' detector geometry, where the optic axis of the SEM intersects the centre of the phosphorous screen, allows acquisitions of Kikuchi patterns directly below the sample where the signal yield is strongest and gnomonic distortions are minimised. This improves the acquisition speed, lateral spatial resolution and indexing rates without sacrificing angular resolution [6-8]. Automated crystal orientation mapping in the transmission electron microscope (ASTAR) is capable of achieving a lateral resolution of 1 nm, similar to 'on-axis' TKD, but overlapping of fine grains in the sample thickness will result in the formation of composite diffraction patterns which can lead to deconvolution problems during the template matching process. We will show in the paper that this is less of a problem in 'on-axis' TKD because of the region in the sample where the Kikuchi patterns are generated. In addition, dynamical errors are always present in ASTAR analysis, and can be a major factor limiting angular resolution [9]. These problems can be addressed by the application of 'on-axis' TKD, and make this technique particularly suitable for the study of the phase distribution and crystallography of complex nano-scale materials.

A better understanding of the spatial resolution, angular resolution and Kikuchi pattern formation mechanisms in TKD are still required to understand the potential and the limitations of the technique and so define how it can best be applied to addressing real materials questions at the nano-scale. A number of authors [5, 7, 10-12] have studied the spatial resolution of TKD analysis; especially lateral resolution where it has been shown that combining high accelerating voltages and low sample thicknesses give the best results [5]. By comparison the depth resolution ($d_d$) has been much less investigated. In the same way as two definitions exist for lateral resolution, the physical and effective lateral resolution [10, 13], depth resolution in TKD can also have two definitions. The physical depth resolution ($d_{phy}$) is the maximum depth from which diffraction information is collected and the effective depth resolution ($d_{eff}$) is a measure of how accurately the predominant orientation can be extracted from overlapping patterns from through thickness variations in phase or orientation. By using bi-crystal samples, Brodu et al. [14] and Sneddon et al. [15] have experimentally measured the physical depth resolution of Al, Si, Cu and Pt layers exposed to 30 keV electrons based on the visibility of Kikuchi bands originated from the materials closest to the incident electron beam. The physical depth resolution is seen to be both highly materials- and energy-dependent, but rather independent of sample thickness. Based on their results, Brodu et al. [14] proposed the function describing the physical depth resolution to be $d_{phy}=0.026E/\mu_0$, where E is the primary beam energy and $\mu_0$ (nm$^{-1}$) is the absorption coefficient for the thermally diffuse scattered electrons for samples exposed to 100 keV electrons. The practical application of this relationship may be limited due to the availability of data on $\mu_0$ values in a wider range of engineering materials. The experimental determination of the effective depth resolution in 'on-axis' TKD has not yet been reported. Clearly, this effective depth resolution depends both on the physical resolution and the capacity of the software to extract the predominant orientation from overlapping patterns [16, 17]. In conventional EBSD, the values of effective resolution are found to be approximately 3 times smaller than the physical resolution [18], but it is not clear whether this is also true for TKD. In terms of practical applications, a better understanding of both the physical and the effective depth resolutions can provide a guide for preparing samples of ideal thickness for TKD analysis.

In this paper we report on experimental measurements of the physical and effective depth resolutions from an SEM with 'on-axis' TKD system, showing that TKD analysis to be extremely surface sensitive, which needs to be carefully considered when conducting orientation mapping on nano-crystalline materials because different grain structures are analysed when opposite sides of a nano-crystalline foil are scanned. Based on these results, the formation mechanisms of Kikuchi patterns in TKD are also discussed.

## 2. Materials and methods

Electron transparent thin foils were prepared from a Zr-1.0Nb alloy, provided by Westinghouse, using the in-situ FIB lift-out method on a Zeiss Crossbeam 540 FIB/SEM system. Following the steps described in [19], lamellar specimens were lifted out and welded to a 3 mm Cu grid using an in-situ micro-manipulator. Progressive thinning with a gradually reducing milling current of 1500-100 pA at 30 kV was performed on both sides of the foil. Final thinning to electron transparency and cleaning was performed at 5 kV and 200 pA. The TKD study was carried out on a Zeiss Merlin FEG-SEM system equipped with a Bruker e-flash high-resolution EBSD detector and an OPTIMUS™ TKD head, Fig. 1. The detailed mapping parameters used are summarized in Table 1. The pattern background correction and indexing was performed by ESPRIT 2.0 software from Bruker. The crystallographic information for each phase used in identifying the phase distribution in the sample is summarized in Table 2. The thicknesses of the regions of interest in the samples studied were determined using Electron Energy Loss Spectroscopy (EELS) in a JEOL ARM 200F microscope operating at 200 kV with a step size of 2 nm, a convergence angle of 30 mrad and a collection angle of 40 mrad using the $t/\lambda$ method suggested by Malis et. al. [20].

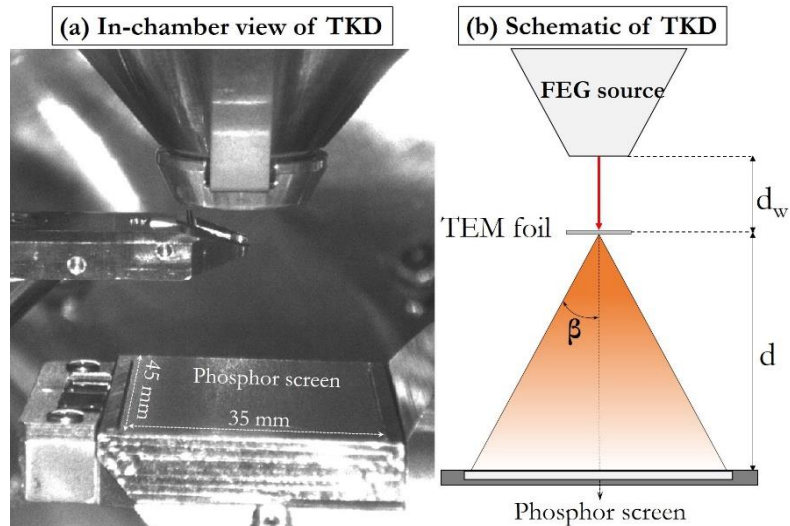

Fig. 1 In-chamber camera view and schematic of on-axis TKD setup

Table 1 Summary of TKD settings

| HT (kV) | Probe current (nA) | Working distance, $d_w$ (mm) | Detector distance, d (mm) | Step size (nm) | Dwell time (ms) | Pattern resolution |
|---|---|---|---|---|---|---|
| 30 | 1.5 | 5 | 16.5 | 4 | 9.3 | 320x240 |

Table 2 Crystal structures used for the TKD analysis.

| Crystal structure | Space group | a(Å) | b(Å) | c(Å) | α(°) | β(°) | γ(°) | Ref |
|---|---|---|---|---|---|---|---|---|
| α-Zr | P63/mmc | 3.23 | 3.23 | 5.14 | 90 | 90 | 120 | [21] |
| β-Nb | Im$\bar{3}$m | 3.3 | 3.3 | 3.3 | 90 | 90 | 90 | [21] |

These particular Zr alloy samples were selected for study because they have very well defined microstructure containing second phase particles (SPPs) of a convenient size for these experiments. The final annealing step of the Zr-Nb sample at 560 °C leads to the formation of recrystallized α-Zr grains with the Nb mostly in solid solution and small Nb-containing particles that are generally spherical with an average diameter ~50 nm [22]. With a typical sample thickness of ~80 nm, some particles will be fully embedded inside the foil while the others may lie closer to one surface than the other and indeed may be truncated by the foil surface. TKD orientation mapping was carried out for the same region but from both sides of the TEM

Page | 3

foil using the same mapping parameters. The physical depth resolution can thus be estimated by the visibility of Kikuchi bands in carefully selected regions with overlapping phases, and the effective resolution estimated depending on the capability of the software to deconvolute the predominant orientation from the overlapping Kikuchi patterns. These same Zr-Nb alloys are being studied as corrosion-resistant nuclear fuel cladding materials [2], and the oxides formed by aqueous corrosion have a characteristic nanoscale grain structure [23, 24] which can also be studied by TKD analysis.

## 3. Results

Fig. 2 shows bright field (BF) and dark field (DF) forescattered electron (FSD) images and the corresponding TKD maps acquired from the same region on the sample but scanned from opposite sides (the sample was flipped on the holder in between acquisitions). The FSD images are generated by collecting the forescattered electrons distributed within different angular ranges. As with scanning transmission electron microscopy (STEM) [25], the low-angle forescattered (LAFSD) image shows more diffraction contrast, Fig. 2 (a), while the high-angle forescattered (HAFSD) image shows more Z (atomic number) contrast, Fig. 2 (b). More detail about FSD imaging can be found in [26]. As the signal contributing to the formation either LAFSD or HAFSD images is integrated from the whole thickness of the specimen, all the particles whether embedded fully or partially in the thin foil can be detected in these FSD images. In comparison with the FSD images, both the number density and the size of particles detected by the TKD band contrast (BC) images and the phase maps are smaller, Fig. 2 (c-f). Of the ten particles observed in the FSD images, only one can be seen in both of the TKD maps scanned in opposite-directions, highlighted by the yellow rectangles in Fig. 2 (a, c and e). We can also see that the apparent size of this particle in both of the TKD BC maps is smaller than in the FSD images. Two other particles are visible in only one of the TKD maps, highlighted by blue and green rectangles in Fig. 2 (a, c and e), indicating these particles are embedded near one surface of the thin foil. All the other particles, highlighted by the red arrows in Fig. 2 (b), are invisible in both of the TKD BC maps regardless of the particle size. The hydrides, which are artefacts of the FIB sample preparation [27], also show different morphology through the thin foil thickness, Fig. 2 (d and f).

However if we look at the FSD images and TKD BC maps carefully, in the regions with particles fully embedded, although they are not resolved in the phase maps we still can see a contrast/brightness difference from those regions compared with the surrounding matrix, e.g. the regions highlighted by black dashed arrows in Fig. 2 (c and e). The contrast of a TKD BC map is generally related to the image quality of the captured patterns, essentially based on the contrast (or sharpness) of the Kikuchi bands compared to the background, and this contrast is useful in giving a visual rendering of features that cause poor band contrast [28], e.g. grain boundaries, phase interfaces and surface imperfections. In our case, the poor band contrast will probably be a result of the complex overlapping patterns caused by the analysis of buried SPPs in a matrix. Depending on the size of the particle and the local thickness of the matrix above and below it, the contrast changes caused by the overlapping phases differ from region to region. In Fig. 2, particles of diameter smaller than 20 nm that are either fully or partially embedded in the foil show little influence on the band contrast if the mapped region has a thickness between 55 nm and 75 nm. The critical value of particle size that appears to induce detectable changes in the band contrast is observed to be ~ 50 nm in diameter in a foil of total thickness of ~70 nm; an example is highlighted by blue dashed arrows in Fig. 2 (c and e).

Apart from a poorer band contrast, an embedded particle can also affect the pattern indexing reliability/accuracy of TKD data from the underlying matrix. This can be reflected in the mean angular deviation (MAD) distribution shown in Fig. 3. The MAD indicates the misfit between the measured and the calculated angles between bands, so the larger the MAD value, the higher the misfit and the lower the indexing confidence. Data from the region covering the particle shows a larger mean value of MAD when compared with that from the surrounding region, as a result of both low pattern quality caused by overlapping patterns and also the existence of lattice strain around the buried particle.

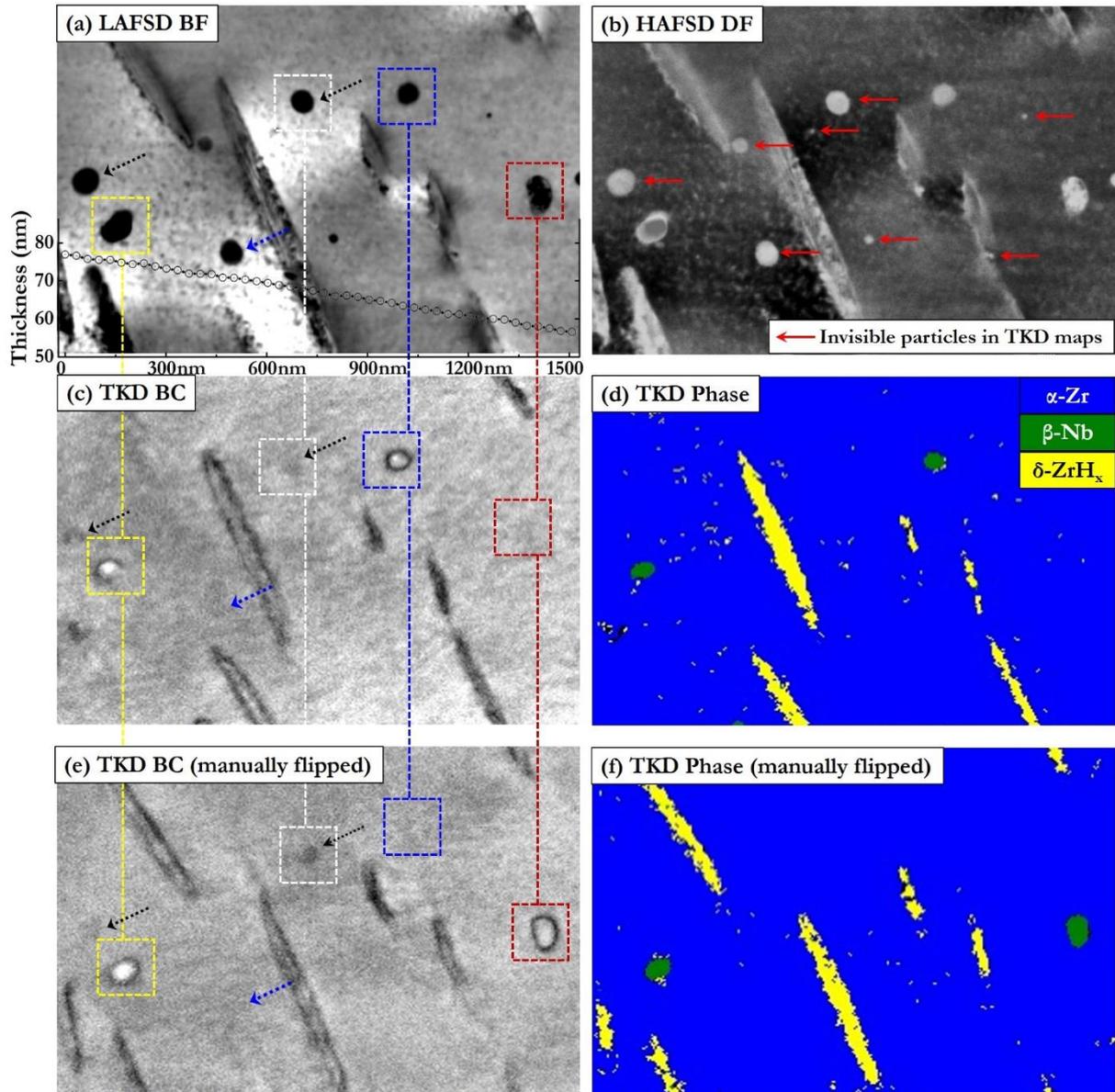

Fig. 2 (a, b) BF and DF images using forescattered electrons and showing the size and distribution of Nb particles. (c, d, e and f) TKD band contrast (BC) and phase maps from scanning the same region of the Zr-Nb sample in both directions. The corresponding thickness profile of the mapped region is plotted in (a).

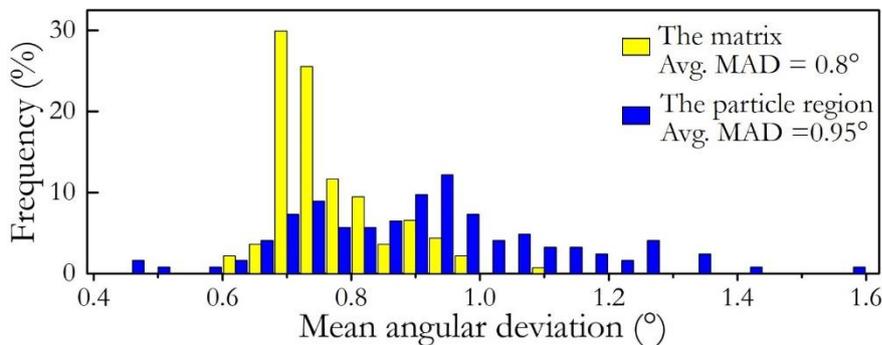

Fig. 3 Frequency distribution of mean angular deviation (MAD) from the region highlighted by the white rectangle in Fig. 2 (e) (blue bars), and from the surrounding matrix (yellow bars) of a same-size region.

Fig. 4 shows the HAFSD image and TKD phase map from one α-Zr grain containing several β-Nb particles. All the particles are seen in the HAFSD image, but only those near the bottom surface of the foil in the TKD map. From these images, we can identify 3 particles with different sizes and different locations

Page | 5

through the foil thickness, labelled P1, P2 and P3, as shown in the schematic diagram in Fig. 4(c) which also captures the measured foil thickness variations in this region of the sample. Kikuchi patterns acquired from the regions containing each of these particles and the matrix are also shown in Fig. 4, with the Kikuchi bands originating from the α-Zr matrix highlighted by yellow dashed lines and from β-Nb by red dashed lines. When particles are only partially embedded in the TEM foil, we have assumed that they are spherical and that roughly half of the particle remains inside the foil, P3 in Fig 4(c), such that the maximum depth that they penetrate into the foil surface is the observed radius. In the region of P3, the dominant bands are diffracted from the β-Nb phase, Fig. 4(g), while bands from the α-Zr matrix show only very weak contrast. This indicates that the depth of penetration of particle P3 up into the foil, ~20 nm, is close to the physical depth resolution of TKD information under these specific imaging and sample conditions. From the HAFSD and TKD images, and Fig. 2, we can conclude that particles P1 (diameter ~60 nm) and P2 (diameter ~50 nm) are both fully embedded in the TEM foil. In the pattern acquired from the region of P2, Fig. 4(f), the α-Zr bands are very well defined but there is little sign of diffraction from the relatively large Nb-rich precipitate, simiar to the pattern acquired for the matrix, Fig. 4(e). In comparison, in the pattern acquired from the region of P1, Fig. 4(d), even though the dominant Kikuchi bands are still indexed as α-Zr, Fig. 4(b and d), some well-defined bands from the β-Nb phase can be seen, Fig. 4(d). If as defined above, we assume that signals contributing to detectable Kikuchi bands originate from a depth of ~20 nm (physical depth resolution) from the bottom surface, if we further assume that particle P1 is in the middle of the thin foil, the thickness of the two phases that contribute to the formation of Kikuchi bands in the region of P1 can be estimated to be ~10 nm α-Zr and 10 nm β-Nb. We thus can estimate that the effective depth resolution for α-Zr to be ~10 nm under these imaging conditions, coresponding to approximately ½ the physical depth resolution.

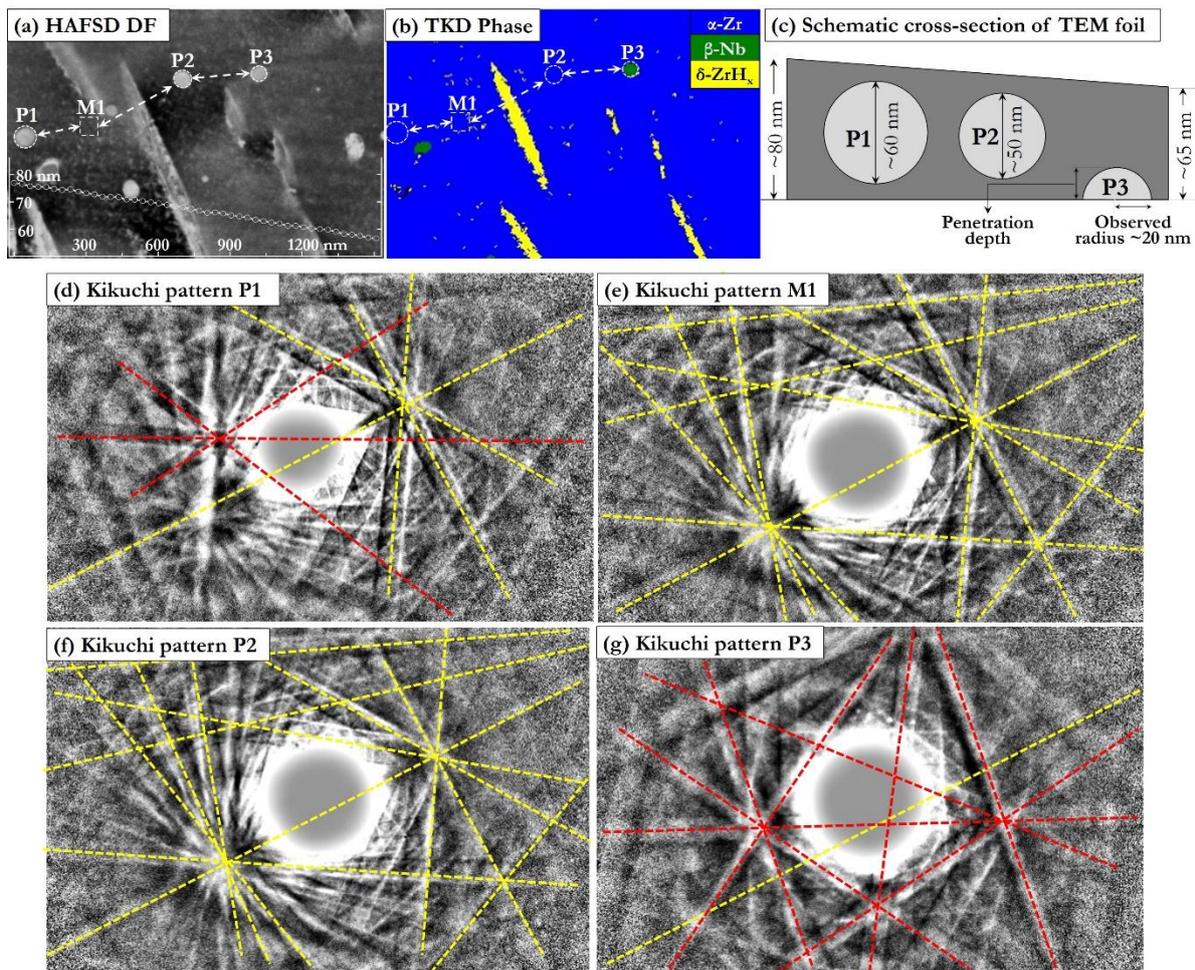

Fig. 4 The influence of overlapping phases on Kikuchi patterns. (a) and (b) are the HAFSD image and TKD phase map showing the locations from which Kikuchi patterns were acquired, and (c) is a schematic

cross-sectional view of the TEM foil showing possible locations of 3 selected particles. Images d, e, f and g show the details of Kikuchi patterns obtained from the locations of these 3 particles and the surrounding matrix. The Kikuchi bands from the β-Nb phase are highlighted by the red dashed lines and α-Zr by yellow dashed lines.

Fig. 5 shows a series of Kikuchi patterns acquired from a similar Zr-Nb sample, but from regions of different thickness. Clear and indexable Kikuchi patterns can be seen from the region of thickness ~120nm, but the intensity of the background is clearly stronger compared with the patterns generated from thin regions above as a result of thermal diffuse scattering (TDS). Discrete diffraction spots start to be seen at a thickness of 50 nm, and become more obvious at 30 nm as expected. The average values of MAD are also shown in Fig. 5. Under the same imaging conditions, these MAD values increase gradually with the sample thickness from 30 nm to 120 nm indicating a gradually decreasing indexing confidence, which is mainly caused by the variation in sample thickness and on diffuse scattering but also could be a result of local strain [17, 29].

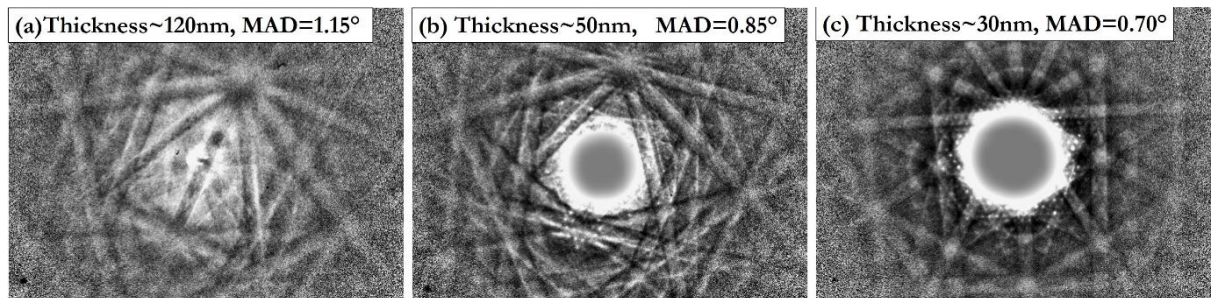

Fig. 5 Examples of Kikuchi patterns acquired from a Zr-Nb sample from regions of different thickness

## 4. Discussion

When a thin foil with embedded particles is scanned in TKD, depending on the combination of microscope parameters such as accelerating voltage, and materials properties like density and atomic mass, different volume of material above the bottom surface will contribute strongly to the formation of the detected Kikuchi patterns. Depending on the capabilities of the indexing software, mounting the sample the other way up may give different apparent microstructures; especially the particle size and number density. Based on the experimental results presented above, Fig. 6 shows a schematic illustrating our suggested correlation between the buried depth of β-Nb particles in a thin foil and their visibility in TKD. It is also clear that different microstructures will be seen by different techniques even from the same sample. For example, by collecting signals integrated from the whole foil thickness, e.g. TEM or FSD imaging, we can detect all the particles shown in Fig. 6. However, the particles that are close to the top surface, e.g. the ones labelled 3, 4 and 5 in Fig. 6, are the only ones shown by secondary electron imaging, with an information depth of a few nanometers [30]. By comparison, in a TKD map only the particles that are distributed close to the bottom surface and have a size larger than the effective depth resolution can be detected. The existing surface-sensitive nature of TKD [14, 31] thus needs to be taken into consideration, especially when analysing nano-crystalline materials with grain size of the order of the sample thickness. An example of a nanocrystalline sample of zirconium oxides scanned by TKD over the same region but from opposite directions is shown in Fig. 7. It can be seen that from foil ~100 nm thick of a material with a grain size that ranges from tens to hundreds of nanometers mounting the sample the other way up results in images with significantly different grain structures.

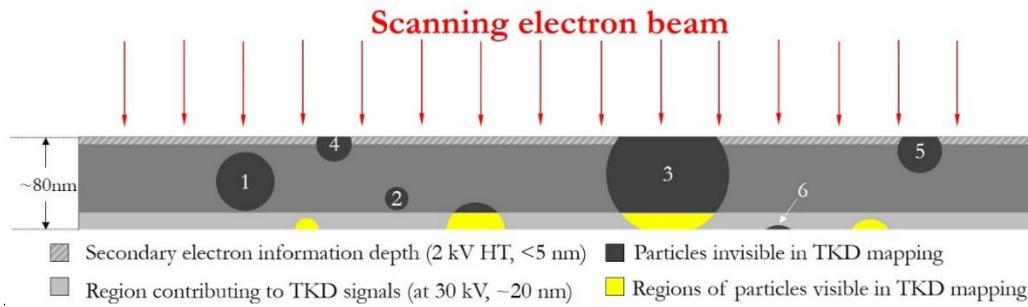

Fig. 6 Schematic of the correlation between the depth of Nb particles in an electron transparent Zr thin foil and their visibility in TKD mapping at 30 kV

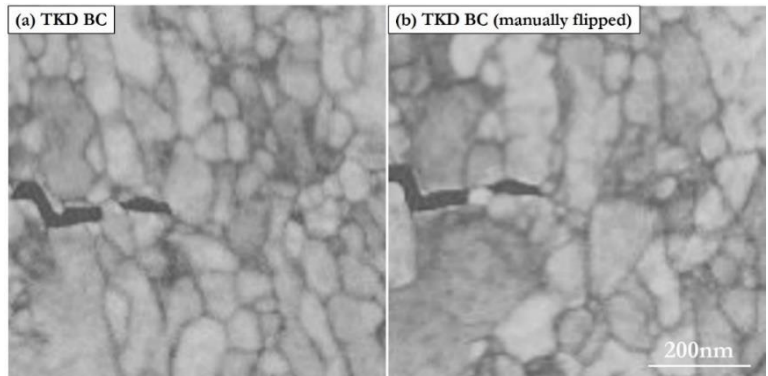

Fig. 7 TKD band contrast maps scanned from opposite sides of the same region in a nano-crystalline zirconium oxide showing different grain structures

Before we can propose a model for the depth resolution, an understanding of the formation mechanisms of Kikuchi patterns in TKD is required. Discrete diffraction spots, diffraction lines/bands or a combination of both can be observed under different diffraction conditions. In TEM, discrete diffraction spots arise from coherent scattering of the incident beam, while the formation of Kikuchi bands is described as a two-step process consisting incoherent scattering of the primary beam followed by coherent scattering of these forward biased electrons [13, 32-34], Fig. 8. Several papers [35-38] discuss the thermal diffuse scattering (TDS) caused by thermal vibrations of the atoms as a major mechanism for incoherent scattering, and every such incoherent scattering event then acts as a point source of electrons that can contribute either to the formation of Kikuchi bands or the background intensity [35, 38, 39], Fig. 8. As electrons generated by TDS contain no crystallographic information, they can lead to a diffuse background intensity that suppresses the contrast and sharpness of Kikuchi bands, especially when TDS arises close to the bottom surface of the foil [38, 40], as highlighted in blue in Fig. 8.

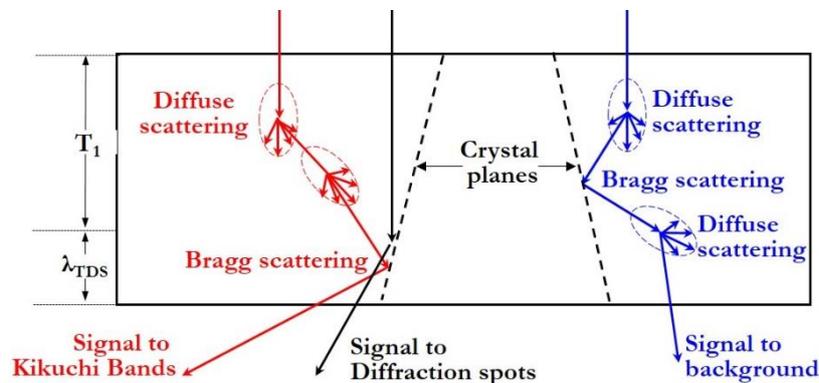

Fig. 8 Schematic diagrams showing multiple scattering processes in a crystal contributing to the formation of Kikuchi bands (red), diffraction spots (black) and the background (blue). The fan-shaped diffuse scattering represents the intensity angular distribution of the TDS electrons.

The physical depth resolution is the depth from which Kikuchi patterns originate in the on-axis TKD geometry used here. Based on the formation mechanisms of Kikuchi bands described above, it is reasonable to infer that the physical depth resolution ($d_{phy}$) of Kikuchi patterns formed in TKD is related to the position of the deepest thermal diffuse scattering event contributing to the pattern, or the TDS mean free path ($\lambda_{TDS}$), $d_{phy} \approx \lambda_{TDS}$. However, only a few experimental data on physical depth resolution can be found in the literature, and even less data on values of $\lambda_{TDS}$, especially for the low energy range of 10 to 30 keV. In Table 3, we summarize all the values of TKD physical depth resolution we could find, and the mean free paths of the corresponding scattering events. Inelastic scattering is a very general term which refers to any process which causes the primary electron to lose a detectable amount of energy, including plasmon scattering, single valence electron excitation and inner shell excitation. The inelastic mean free paths ($\lambda_{IMFP}$) for 41 elemental solids for electron energies from 50 eV to 30 keV can be found in [41]. TDS is not fully incoherent but also includes an inelastic component with an average energy loss per event less than 0.1 eV [42]. However, values of $\lambda_{TDS}$ for 30 keV primary beam are not currently available, and those shown in Table 3 are calculated based on a 100 keV primary beam. It can be seen that values of physical depth resolution correlate better with $\lambda_{TDS}$ rather than $\lambda_{IMFP}$, and these slight differences might be further reduced if more accurate low voltage data on $\lambda_{TDS}$ were available. The predicted correlation between $\lambda_{TDS}$ and physical depth resolution is also in agreement with Monte-Carlo simulations carried by van Bremen et al. [5] in which they calculated the depth at which the final incoherent electron scattering events occurred, and reported a physical depth resolution of ~12 nm in Au samples (Z=79, $\varrho$=19.3 g.cm$^{-3}$, $\lambda_{IMFP}$= 21.3 nm and $\lambda_{MFP}$= 3.5 nm) exposed to 30 keV electrons, a material in which we would expect there to be a similar depth resolution as experimentally determined in Pt, 13 nm [15]. A similar relationship has been reported by Brodu et al. [14] for Si samples exposed to 30 keV electrons, $d_{phy} \approx 0.026E/\mu_0$ (where E is the beam energy and $\mu_0$ is the absorption coefficient for 100 keV electrons in Si). If we consider the relationship between $\lambda_{TDS}$ and absorption coefficient to be $\lambda_{TDS} \approx 1/\mu_0$ [42], the Brodu relationship is then similar to the value of $d_{phy}$ proposed above.

Table 3 Summary of experimentally determined values of TKD physical depth resolution at 30 keV and the mean free paths of scattering events in the corresponding materials

|  | Al | Si | Cu | Nb | Pt |
|---|---|---|---|---|---|
| Physical depth resolution, $d_{phy}$, (nm) | 80 [15] | 60 [14] | 33 [15] | 20 | 13 [15] |
| Inelastic scattering mean free path, $\lambda_{IMFP}$, (nm) [41] | 37.6 | 43 | 27 | 30.8 | 20.5 |
| Elastic scattering mean free path, $\lambda_{MFP}$, (nm) [43] | 23.7 | 24.8 | 8.0 | 5.7 | 3.5 |
| TDS mean free path[1], $\lambda_{TDS}$, (nm) | 90 | 87 | 40 | - | - |
| Ratios ($d_p/\lambda_{TDS}$) | 0.9 | 0.7 | 0.8 | - | - |
| Ratios ($d_p/\lambda_{MFP}$) | 3.4 | 2.4 | 4.1 | 3.5 | 3.7 |
| Ratios ($d_p/\lambda_{IMFP}$) | 2.1 | 1.4 | 1.2 | 0.6 | 0.6 |
| Ratios ($d_p/\varrho$) | 29.6 | 25.9 | 3.7 | 2.3 | 0.6 |
| Ratios ($d_p/Z$) | 6.2 | 4.3 | 1.1 | 0.5 | 0.2 |
| TDS absorption coefficient[2], $\mu_0$, (nm$^{-1}$) [44] | 0.6 | 0.6 | 1.4 | - | - |
| Atomic Number, Z | 13 | 14 | 29 | 41 | 78 |
| Density, $\varrho$, (g.cm$^{-3}$) | 2.7 | 2.3 | 9.0 | 8.6 | 21.5 |

[1] The values of $\lambda_{TDS}$ are estimated using equation, $\lambda_{TDS} \approx 1/\mu_0$ [42]
[2] The values of absorption coefficient are calculated for 100 keV primary electrons

In addition, the ratio between the physical depth resolution and the elastic scattering mean free path ($\lambda_{MFP}$) is nearly constant, Table 3, indicating that the diffracted intensity contributing to the Kikuchi bands generated inside the sample has a high probability to escape the sample surface if the TDS position is within a distance < $3.5\lambda_{MFP}$ or < $\lambda_{TDS}$ from the bottom surface. The implicit relationship that $\lambda_{TDS} \approx 3.5\lambda_{MFP}$ will need more calculations to confirm, and this lies beyond the scope of the current study. The outlier to this relationship is the data for Si, Table 3, where the determination of the physical depth resolution is based on the visibility of Kikuchi bands originated from the top crystal closest to the incident electron beam in a bi-crystal Si sample [14]. However, the authors did not take the influence of the thickness of the upper grain in their bi-crystal sample into consideration, which can affect the intensity of Kikuchi bands associated with this crystal. At the critical thickness of the bottom crystal, ~60 nm [14], which they regard as the physical resolution, the thickness of the top crystal is only ~40 nm, smaller than $\lambda_{TDS}$, and could limit the intensity of Kikuchi bands generated. It is thus reasonable to infer the physical depth resolution for Si exposed to 30 keV electrons could be larger than 60 nm. If we assume that a value of 80 nm to be more realistic, then the ratio between physical depth resolution and $\lambda_{MFP}$ in all the materials reported is close to be 3.5.

Many papers have shown that sample thickness is very important in TKD analysis [12, 45]. By considering the role of TDS on the formation of Kikuchi bands, and so on the physical resolution and the intensity of the background as discussed above, we may thus suggest a suitable sample thickness to achieve clear and indexable Kikuchi patterns for TKD analysis to be $\leq 6\lambda_{TDS}$, or $21\lambda_{MFP}$ for samples of high crystalline symmetry. As the observed band intensity is proportional to the square of the structure factor amplitude [13, 46], we can expect the contrast and sharpness of Kikuchi bands in samples of lower crystalline symmetry will be more sensitive to the sample thickness, and samples $\leq 3\lambda_{TDS}$, or $10\lambda_{MFP}$ are suggested.

## 5. Conclusions

The formation mechanisms of Kikuchi bands in TKD are believed to follow a two-step process similar to that in TEM; incident electrons first undergoing incoherent thermal diffuse scattering followed by Bragg scattering from crystal planes. The physical depth resolution of TKD is then logically related to the depth where the deepest TDS event occurs. A function describing the physical depth resolution is thus suggested to be $d_{phy} \approx \lambda_{TDS}$, the mean free path of TDS in materials exposed to electrons of specific primary energies. The value of effective depth resolution is found to be approximately half of the physical depth resolution. Based on the observed ratios between the physical depth resolution and the mean free path of elastic scattering, we propose that $d_{phy}$ can be estimated as $3.5\lambda_{MFP}$. Compared with the limited data available for $\lambda_{TDS}$, especially in the lower energy range from 10 keV to 30 keV used in TKD analysis, $\lambda_{MFP}$ for electrons with kinetic energies from 10 keV to 30 keV and materials with atomic number Z= 1 to 92 can be easily calculated using the elastic scattering cross sections data in [43]. We suggest a guideline for the sample thickness $\leq 6\lambda_{TDS}$, or $21\lambda_{MFP}$ for samples of high crystalline symmetry, and lower values for samples of lower symmetry, in order to achieve clear and indexable Kikuchi patterns for TKD analysis.

### Acknowledgements

The authors acknowledge the MUZIC project for providing zirconium samples. EPSRC grants (EP/K040375/1 and EP/N010868/1) are acknowledged for funding the 'South of England Analytical Electron Microscope' and the Zeiss Crossbeam FIB/SEM used in this research.